\documentclass[final,5p,times,twocolumn]{elsarticle}

\usepackage{graphicx}
\usepackage{amssymb}
\usepackage{amsmath}
\usepackage{amsthm}
\usepackage[normalem]{ulem}
\biboptions{sort&compress}
\usepackage{slashed}
\usepackage{color,rotating}
\usepackage{bbm}
\usepackage{array}
\usepackage[utf8]{inputenc}
\usepackage{subcaption}
\usepackage{multirow}
\usepackage{xcolor}

\allowdisplaybreaks

\begin{document}

\newcommand{\p}{\partial}
\newcommand{\hh}{{\widehat{h}}}
\newcommand{\bchi}{{\bar{\chi}}}
\newcommand{\btheta}{{\bar{\theta}}}
\newcommand{\ds}{{\slashed{\nabla}}}
\newcommand{\unit}{{\mathbbm{1}}}

\newcommand{\gb}{\bar{g}}
\newcommand{\Db}{\bar{D}}
\newcommand{\Rb}{\bar{R}}

\newcommand{\cO}{\mathcal{O}}
\newcommand{\cR}{\mathcal{R}}

\newcommand{\be}{\begin{equation}}
\newcommand{\ee}{\end{equation}}

\begin{frontmatter}

\title{Asymptotically Safe Gravity with Fermions}

\author{Jesse Daas}
\ead{j.daas@science.ru.nl}
\author{Wouter Oosters}
\ead{w.oosters@student.ru.nl}
\author{Frank Saueressig}
\ead{f.saueressig@science.ru.nl}
\author{Jian Wang}
\ead{jian.wang@science.ru.nl}

\address{Institute for Mathematics, Astrophysics and Particle Physics (IMAPP) \\ Radboud University, Heyendaalseweg 135, 6525 AJ Nijmegen,The Netherlands \\[2ex]
This article is registered under preprint number: arXiv: 2005.12356
}

\begin{abstract}
We use the functional renormalization group equation for the effective average action to study the fixed point structure of gravity-fermion systems on a curved background spacetime. We approximate the effective average action by the Einstein-Hilbert action supplemented by a fermion kinetic term and a coupling of the fermion bilinears to the spacetime curvature. The latter interaction is singled out based on a ``smart truncation building principle''. The resulting renormalization group flow possesses two families of interacting renormalization group fixed points extending to any number of fermions. The first family exhibits an upper bound on the number of fermions for which the fixed points could provide a phenomenologically interesting high-energy completion via the asymptotic safety mechanism. The second family comes without such a bound. The inclusion of the non-minimal gravity-matter interaction is crucial for discriminating the two families. Our work also clarifies the origin of the strong regulator-dependence of the fixed point structure reported in earlier literature and we comment on the relation of our findings to studies of the same system based on a vertex expansion of the effective average action around a flat background spacetime.
\end{abstract}


\end{frontmatter}

\section{Introduction}
\label{sect.1}
Any realistic quantum theory for the gravitational interactions has to incorporate matter degrees of freedom in one way or another. Minimalistically, one could opt for a matter sector comprising the field content of the standard model of particle physics but extensions by additional fields are phenomenologically interesting options as well. From a quantum gravity viewpoint, it is then conceivable that consistency of the theory constrains the admissible matter sectors. A prototypical example is string theory where supersymmetry dictates that every bosonic degree of freedom has to be paired with a fermionic partner.

For the gravitational asymptotic safety program \cite{Percacci:2017fkn,Reuter:2019byg},\footnote{Also see  \cite{Christiansen:2014raa,Christiansen:2015rva,Codello:2015oqa,Gies:2015tca,Gies:2016con,Ohta:2016npm,Ohta:2016jvw,Denz:2016qks,Labus:2016lkh,Knorr:2017mhu,Christiansen:2017bsy,Lippoldt:2018wvi,Morris:2018zgy,Bosma:2019aiu,Knorr:2019atm,Falls:2020tmj} for recent developments and \cite{Bonanno:2020bil} for a detailed bibliography.} reviewed in \cite{Niedermaier:2006wt,Codello:2008vh,Litim:2011cp,Reuter:2012id, Ashtekar:2014kba,Eichhorn:2018yfc,Reichert:2020mja}, the addition of matter fields is conceptually straightforward. The program lives on the so-called theory space comprising all actions which can be constructed from a given field content and are compatible with the postulated symmetry requirements. Gravity-matter theories can then be studied by supplementing the gravitational degrees of freedom by  matter fields and extending the set of actions spanning the theory space. The asymptotic safety condition then restricts the admissible matter sectors by requiring the existence of a non-Gaussian renormalization group fixed point (NGFP) which could provide the high-energy completion of the theory.

As recently reviewed in \cite{Eichhorn:2018yfc}, there has been significant effort towards understanding the role of asymptotic safety in gravity-matter systems \cite{Benedetti:2009rx,Benedetti:2009gn,Manrique:2010mq,Eichhorn:2011pc,Dona:2013qba,Becker:2014jua,Oda:2015sma,Dona:2015tnf,Eichhorn:2016esv,Biemans:2017zca,Christiansen:2017cxa,Hamada:2017rvn,Eichhorn:2017eht,Eichhorn:2017sok,Bonanno:2018gck,Grabowski:2018fjj,Alkofer:2018fxj,Alkofer:2018baq,Maas:2019eux,Eichhorn:2019dhg,deBrito:2019epw,Reichert:2019car,Ohta:2020bsc}. This led to some remarkable insights. Firstly, asymptotic safety puts only mild conditions on the admissible matter sectors \cite{Dona:2013qba,Biemans:2017zca,Alkofer:2018fxj}. In particular the matter content of the standard model may give rise to a NGFP suitable for rendering the theory asymptotically safe. Secondly, NGFPs associated with gravity-matter systems can exhibit an enhanced predictive power as compared to the asymptotically free theory. Quantum fluctuations present at the NGFP can turn a power-counting marginal coupling into an irrelevant one thereby fixing some of the free parameters. Examples where such a mechanism may be operative include the Higgs mass \cite{Shaposhnikov:2009pv,Eichhorn:2017als,Pawlowski:2018ixd,Kwapisz:2019wrl}, the fine-structure constant \cite{Harst:2011zx,Eichhorn:2017lry}, or ratios among quark masses \cite{Eichhorn:2018whv}. 

 A crucial ingredient in developing phenomenologically interesting gravity-matter theories within the asymptotic safety program is the inclusion of fermionic degrees of freedom. Based on the Wetterich equation the existence of suitable NGFPs has been investigated within background field computations \cite{Dona:2012am,Hamada:2017rvn,Biemans:2017zca,Alkofer:2018fxj,Alkofer:2018baq}, the fluctuation field approach \cite{Meibohm:2015twa,Meibohm:2016mkp,Eichhorn:2018ydy,Eichhorn:2018nda}\footnote{For further results based on the fluctuation field approach also see \cite{Christiansen:2012rx,Codello:2013fpa,Christiansen:2014raa,Christiansen:2017cxa,Eichhorn:2018akn,Burger:2019upn} and the review \cite{Pawlowski:2020qer}.}, and settings employing a hybrid of these two computational strategies \cite{Dona:2013qba,Dona:2015tnf}. In particular, non-minimal couplings of the fermionic degrees of freedom to gravity have been considered in \cite{Eichhorn:2016vvy,Eichhorn:2018nda}. An intriguing property of the background computations is the potential presence of a (regularization-procedure dependent) upper critical value on the number of fermions which, once exceeded, could impede the phenomenological viability of the corresponding NGFP. This situation is rather unsatisfactory, since the number of fermions contained in the standard model of particle physics, $N_f^{\rm SM} = 22.5$, exceeds typical values for $N_f^{\rm crit}$ and only the inclusion of gauge fields renders the model suitable for a high-energy completion via the asymptotic safety mechanism \cite{Dona:2013qba}. The goal of this work is to provide a detailed analysis of this situation. In this course, we identify and characterize a new one-parameter family of gravity-fermion fixed points which are invisible in approximations where the fermions are minimally coupled.

 Starting from the setting \cite{Dona:2012am}, comprising the Einstein-Hilbert action supplemented by minimally coupled Dirac fermions, we identify a specific fermion-spacetime curvature interaction which is crucial for understanding the fixed point structure of the system in sect. \ref{sect.2}. The extended truncation is analyzed in sect.\ \ref{sect.3} where we show that the system actually possesses two infinite families of NGFPs. The first one has been identified in \cite{Dona:2012am} and is located at a point where physical properties are particularly sensitive to the regularization procedure. The new family of NGFPs are viable candidates for asymptotically safe gravity-matter systems including an arbitrary number of fermions. Our computation employs the spin-base formalism developed in \cite{Gies:2013noa,Lippoldt:2015cea} and reviewed in \cite{Lippoldt:2016ayw}.
  
\section{Fermions and the Functional Renormalization Group}
\label{sect.2}
We start by reviewing the functional renormalization group equation for gravity coupled to $N_f$ Dirac fermions. As key result we identify the extension \eqref{fermextra} as a canonical candidate for stabilizing NGFPs in gravity-fermion systems.
\subsection{General Setup}
\label{sect.2a}
The key ingredient underlying the asymptotic safety mechanism is a NGFP of the theory's renormalization group (RG) flow. At such a point the theory exhibits an enhanced symmetry, so-called quantum scale invariance \cite{Wetterich:2019qzx}. An RG trajectory whose high-energy behavior is controlled by a NGFP is free from unphysical ultraviolet (UV) divergences and termed ``asymptotically safe'' \cite{Weinberg:1976xy,Weinberg:1980gg}. By definition, these trajectories span the UV-critical hypersurface $\mathcal{S}^{\rm UV}$ of the fixed point. Typically, not all RG trajectories in the vicinity of a NGFP are within $\mathcal{S}^{\rm UV}$ as some may be repelled along an unstable direction. Selecting one specific trajectory within $\mathcal{S}^{\rm UV}$ then requires specifying dim($\mathcal{S}^{\rm UV}$) parameters. All other couplings appearing in the action can be expressed in terms of these ``relevant parameters'', see \cite{Codello:2007bd,Benedetti:2009rx} for explicit examples of such relations. 
Requiring that a NGFP is interesting from a phenomenological perspective gives rise to additional constraints. For instance, the attractiveness of gravity dictates that the NGFP must be  situated at a positive value of Newton's coupling since $G$ cannot switch sign along an RG flow \cite{Reuter:2001ag}. Once suitable matter sectors are added additional consistency tests become available \cite{Eichhorn:2018yfc}.

The primary tool for investigating Asymptotic Safety is the functional
renormalization group equation (Wetterich equation) for the effective average action $\Gamma_k$ \cite{Wetterich:1992yh,Morris:1993qb,Reuter:1996cp} formulated for gravity \cite{Reuter:1996cp}
\be\label{FRGE}
k \p_k \Gamma_k = \frac{1}{2} {\rm Tr} \left[ \left( \Gamma_k^{(2)} + \cR_k \right)^{-1} \,  k \p_k \cR_k \right] \, . 
\ee
The Wetterich equation encodes a Wilsonian RG flow in the sense that it captures the change of $\Gamma_k$ when integrating out a shell of quantum fluctuations with momenta $p^2$ approximately equal to the coarse graining scale $k^2$.  The flow of $\Gamma_k$ is driven by the right-hand side of \eqref{FRGE} where $\Gamma_k^{(2)}$ denotes the second functional derivative of $\Gamma_k$ with respect to the fluctuation fields and the trace contains an integral over loop momenta
and a sum over fields. The regulator $\cR_k$ provides a mass term for fluctuations with momenta $p^2 < k^2$ and vanishes for $p^2 \gg k^2$. The interplay between the regulator function appearing in the numerator and denominator then ensures that the right-hand side is finite and peaked at momenta $p^2 \approx k^2$. 

Notably, the derivation of the Wetterich equation does not require the specification of a fundamental action. Structurally, it equips the theory space associated with a given field content with a vector field. Fixed points appear as $k$-stationary points of this vector field. This feature makes the formalism predestined for investigating the existence of (interacting) RG fixed points and their relevant deformations.

\subsection{Fermions minimally coupled to gravity}
\label{sect.2b}
In the following we focus on gravity-fermion systems in a four-dimensional, \emph{Euclidean} spacetime. Our initial ansatz for $\Gamma_k$ follows \cite{Dona:2012am} and takes the form
\begin{equation}\label{Gans}
\Gamma_k = \Gamma_k^{\rm grav}[g] + \Gamma_k^{\rm ferm}[g,\bar{\psi},\psi]
\end{equation}
supplemented by a harmonic gauge fixing condition \cite{Reuter:1996cp} and the corresponding ghost action. Here $g_{\mu\nu}$ and $\psi$ denote the spacetime metric and the Dirac spinors and we suppressed an index enumerating the $\psi$'s. Following \cite{Dona:2012am}, we approximate $\Gamma_k^{\rm grav}$ by the Einstein-Hilbert action,
\be\label{EHaction}
\Gamma_k^{\rm grav}[g] = \frac{1}{16 \pi G_k} \int d^4x \sqrt{g} \left[ - R + 2 \Lambda_k \right] \, ,
\ee
including a scale-dependent Newton's coupling $G_k$ and cosmological constant $\Lambda_k$.\footnote{These ``running couplings'' must not be confused with the renormalized couplings appearing in the effective action which are scale-independent \cite{Knorr:2019atm}.} The fermions are minimally coupled,
\begin{equation}\label{fermion-action}
\Gamma_k^{\rm ferm}[g,\bar{\psi},\psi] = \int d^4x\sqrt{g} \, \bar{\psi} \, \big[ \,i\ds + m \, \gamma^5 \big] \psi \, .
\end{equation}
Here $\ds = \gamma^\mu\nabla_\mu$ is the Dirac operator constructed from the spin-covariant derivative $\nabla_\mu$ and the $\gamma$-matrices satisfy the Clifford algebra $\gamma^\mu\gamma^\nu + \gamma^\nu\gamma^\mu = 2 g^{\mu\nu}$. Furthermore, $\gamma^5$ is fifth gamma-matrix obeying $(\gamma^5)^2 = \unit$. The presence of $\gamma^5$ in the mass term is owed to our conventions for the fermion kinetic term where the relation $\gamma^\mu\gamma^5 + \gamma^5\gamma^\mu = 0$ implies that the square of the Dirac equation gives rise to the Klein-Gordon equation. Moreover, it ensures that the Clifford algebra, the reality conditions obeyed by the Dirac operator, and the Lichnerowicz formula \eqref{Lichnerowiczformula} are mutually consistent in Euclidean signature. In the following, we focus on the case of massless fermions setting $m=0$.

Based on the ansatz \eqref{Gans}, the right-hand side of the Wetterich equation is constructed by considering fluctuations $\{h_{\mu\nu}, \bar{\chi}, \chi \}$ of the fields around a fixed background configuration $\{\gb_{\mu\nu}, \bar{\theta}, \theta \}$. In this work we resort to the linear split
\be\label{background}
g_{\mu\nu} = \gb_{\mu\nu} + h_{\mu\nu}
\, , \quad 
\bar{\psi} = \bar{\theta} + \bar{\chi}
\, , \quad 
\psi = \theta + \chi \, . 
\ee
We then substitute this expansion into the Wetterich equation and read off the scale-dependence of $G_k$ and $\Lambda_k$ at zeroth order in the fluctuation fields. In order to ease the computation we  chose $\gb_{\mu\nu}$ as the metric of the four-sphere and set the background value of the fermions to zero. This suffices to keep track of the two book-keeping (tensor) structures $\int d^4x \sqrt{\gb}$ and $\int d^4x \sqrt{\gb} \Rb$ whose coefficients encode the flow of $G_k$ and $\Lambda_k$. 

The final ingredient in the construction is the regulator function $\cR_k$. In general $\cR_k$ is a matrix valued in field space. In the gravitational and ghost sector the harmonic gauge choice entails that all derivatives contained in $\Gamma_k^{(2)}$ organize themselves into Laplacians $\Delta \equiv - \gb^{\mu\nu} \Db_\mu \Db_\nu$ constructed from the background metric. We then follow \cite{Reuter:1996cp,Codello:2008vh} and construct the entries from $\cR_k$ via the substitution rule (Type I regulator)
\be\label{typeIreg}
\Delta \mapsto P_k(\Delta) = \Delta + R_k(\Delta) \, . 
\ee
The $R_k(\Delta)$ is taken as the Litim regulator \cite{Litim:2000ci,Litim:2001fd}
\be\label{Reg:litim}
R_k = k^2 \, (1-\Delta/k^2) \, \Theta(1-\Delta/k^2) \, , 
\ee  
where $\Theta(x)$ is the unit-step function. Constructing the regulator in the fermionic sector is slightly more involved. Motivated by the mass term appearing in \eqref{fermion-action}, we replace $m$ by a $k$-dependent regulator of dimension one,
\be\label{refferm}
\cR_k^\psi = k \, \gamma^5 \, \left( 1- \sqrt{\Box}/k \right) \, \Theta(1- \sqrt{\Box}/k) \, ,
\ee
where $\Box$ denotes a suitable coarse-graining operator. Following \cite{Dona:2012am}, two canonical choices for the coarse-graining operator are $\Box = - \ds^2$ and $\Box = \Delta$. These choices are related by the Lichnerowicz formula
\be
-\ds^2 = \Delta + \frac{1}{4} \Rb.
\label{Lichnerowiczformula}
\ee
In order to treat both cases simultaneously, we then write
\be\label{boxdef}
\Box = - \ds^2 + \beta \, \Rb
\ee
where $\beta = 0$ and $\beta = -1/4$ correspond to $\Box$ being the (squared) Dirac-operator and the Laplacian, respectively.  The curvature term in \eqref{boxdef} may give a spurious contribution to the flow of the background couplings. The analysis of the system requires an educated guess for this scale and both $\beta = 0$ and $\beta = -1/4$ constitute natural choices. The
properties of the RG flow may be sensitive to the choice of this parameter, in particular, if the approximation of $\Gamma_k$ is taken too simple. The results \cite{Dona:2012am} and the general discussion \cite{Pawlowski:2015mlf} suggest that one has to go beyond the case of minimal coupled matter fields in order to understand the implications of this relative scale in a quantitative way.\footnote{For further discussions on the regulator dependence of truncated renormalization group equations also see \cite{Pawlowski:2005xe,Balog:2019rrg}.}

We start by determining the scale-dependence of $G_k$ and $\Lambda_k$. The explicit computation combines the early-time expansion of the heat-kernel \cite{Codello:2008vh,Vassilevich:2003xt} with standard $\gamma$-matrices manipulations and we refer to the technical companion paper \cite{prc1} for further details. The result is conveniently expressed in terms of the beta functions 
\be\label{betaEH}
\begin{split}
k \p_k \lambda_k = \beta_\lambda(\lambda_k,g_k) 
\, , \quad 
k \p_k g_k = \beta_g(\lambda_k,g_k) 
\end{split}
\ee
for the dimensionless couplings
$g_k \equiv G_k \, k^2$ and 
$\lambda_k \equiv \Lambda_k \, k^{-2}$. 
The explicit form of the beta functions is
\be\label{betas}
\begin{split}
&\beta_g = (2 + \eta_N)g \, \\
&\beta_\lambda = (\eta_N-2)\lambda + \frac{g}{8\pi}\left[ \Big( 20 -\frac{10}{3}\eta_N \Big) \frac{1}{1-2\lambda}-16\right] - \frac{g N_f}{3} \, . 
\end{split}
\ee
The anomalous dimension of Newton's coupling, $\eta_N \equiv (G_k)^{-1} k \p_k G_k$, is conveniently parameterized by
\be
\eta_N = \frac{g \, \big(B_1^{\rm grav}(\lambda) + N_f \, B^{\rm f,minimal} \big)}{1-g \, B_2^{\rm grav}(\lambda)} \, ,
\ee
with 
\be
\begin{split}
 &B_1^{\rm grav} = \frac{1}{3\pi}\left[-\frac{9}{(1-2\lambda)^2} + \frac{5}{1-2\lambda} - 7\right] \, , \\
&B_2^{\rm grav} = \frac{1}{12\pi}\left[\frac{6}{(1-2\lambda)^2} -\, \frac{5}{1-2\lambda} \right] \, , \\
&B^{\rm f,minimal} = -\frac{(\pi-2)}{12\pi} \, \left[ 1 - 12 \, (\beta + \frac{1}{4}) \right]\, .
\end{split}
\ee
The contributions from the fermionic action \eqref{fermion-action} are all proportional to $N_f$ and vanish for $N_f = 0$. The parameter $\beta$ appears in $B^{\rm f,minimal}$ only.

In order to investigate whether the gravity-fermion system admits a high-energy completion 
through the asymptotic safety mechanism, we investigate its fixed points and the stability properties of the flow in their vicinity. At a fixed point $\{u_*^i\}$ all beta functions vanish simultaneously $\beta_{u_j}|_{u^i = u_*^i} = 0$. The flow in the vicinity of such a point is conveniently encoded in the stability matrix
$B_i{}^j = \left. \frac{\p \beta_{u_i}}{\p u_j} \right|_{u^i = u_*^i}$, 
governing the linearized flow equations. 
Introducing the critical exponents $\theta_i$ as minus the eigenvalues of ${\bf B}$, i.e., ${\bf B} V_i = - \theta_i V_i$, every $\theta_i$ with a positive real part is associated with an eigenvector $V_i$ along which the RG flow is dragged into the fixed point as $k$ is increased. Thus the dimension of $\mathcal{S}^{\rm UV}$ (equaling the number of free parameters) is given by the number of $\theta_i$'s with positive real part.

When investigating the beta functions \eqref{betas} for $N_f = 0$ one finds that the pure-gravity system admits a Gaussian fixed point (GFP) at the origin $\{\lambda_*^{\rm GFP}, g_*^{\rm GFP} \} = \{0,0\}$. In addition there is a NGFP situated at $\{\lambda_*^{\rm NGFP}, g_*^{\rm NGFP} \} = \{0.193,0.707\}$. The critical exponents of this fixed point are $\theta_{1,2} = 1.48 \pm 3.04 i$, indicating that the fixed point is UV attractive in both $g_k$ and $\lambda_k$. 

The NGFPs appearing in the minimally coupled gravity-fermion setting are given by the dashed lines in the left panels of Fig.\ \ref{Fig.2}. Notably, one obtains an infinite family of NGFPs, extending to arbitrary values $N_f$, for both $\beta=0$ and $\beta = -1/4$. For $\beta = -1/4$, all NGFPs are located at $g_* > 0$. For $\beta = 0$ there is a critical number of fermions $N_f^{\rm crit} = 12.26$ where the NGFP transitions from $g_* > 0$ ($N_f < N_f^{\rm crit}$) to $g_* < 0$ ($N_f > N_f^{\rm crit}$). Combined with the condition that a phenomenologically viable high-energy completion requires that the NGFP must be situated at $g_* > 0$, one concludes that not all NGFPs may be viable candidates for rendering the gravity-matter system asymptotically safe.  
The finite value of $N_f^{\rm crit}$, appearing for one choice of coarse-graining operator while absent for another, indicates that the validity of asymptotic safety seemingly depends on the choice of regularization procedure. This was the conclusion reached in \cite{Dona:2012am}. Let us stress, however, that the existence of the NGFP is actually independent of the choice of regulator. Merely its position may or may not be suitable for building a viable phenomenology.

\subsection{Smartly extending the effective average action}
\label{sect.2c}
The $\beta$-dependence discussed in the previous subsection suggests that 
 the ansatz for $\Gamma_k$ made in eqs.\ \eqref{EHaction} and \eqref{fermion-action} could be too simple for capturing the essential properties of the NGFPs. In order to improve our approximation systematically, it is useful to understand the mechanism underlying the presence (or absence) of $N_f^{\rm crit}$. In order to facilitate the clarity of the discussion, we will set $\lambda = 0$ and focus on the fixed point equation for Newton's coupling $\eta_N(g_*, \lambda = 0) = - 2$. In this case $g_*$ is determined by the linear equation
\be\label{NGFPred}
g_* (B_1^{\rm grav} - B_2^{\rm grav} + N_f \, B^{\rm f,minimal})|_{\lambda = 0} = -2 
\ee
where $(B_1^{\rm grav} - B_2^{\rm grav})|_{\lambda = 0} = -15/(4\pi) < 0$. This entails that for $N_f = 0$ one has $g_* > 0$. The sign of $B^{\rm f,minimal}$ depends on the choice of $\beta$ though:\footnote{The numerical values of these coefficients differ from the ones reported in \cite{Dona:2012am}. This can be traced back to the different shape of the regulator function \eqref{refferm} employed by the two works. Notably, this has no effect on the qualitative behavior encoded by the signs of the coefficients, showing robustness of this feature under a change of regularization procedure.}
\be
 B^{\rm f,minimal}|_{\beta = -1/4} = \frac{2-\pi}{12 \pi} < 0 \, , \quad
 B^{\rm f,minimal}|_{\beta = 0} = \frac{\pi -2}{6 \pi} > 0 \, .
\ee
If $B^{\rm f,minimal} < 0$ the bracket on the left-hand side of \eqref{NGFPred} is negative for all values of $N_f$ resulting in $g_* > 0$.
If $B^{\rm f,minimal} > 0$ there is a critical value $N_f^{\rm crit}$ where the bracket changes sign and the NGFP transitions from $g_* > 0$ to $g_* < 0$. 
This is the behavior exhibited by the dashed lines in Fig.\ \ref{Fig.2}.

This analysis suggests searching for additional terms in $\Gamma_k$ which contribute to $B^{\rm f,minimal}$. A systematic analysis, on a spherically symmetric background, indeed identifies a canonical interaction term coupling the fermion bilinears to the spacetime curvature
\be\label{fermextra}
\Delta\Gamma_k^{\rm ferm}[g,\bar{\psi},\psi] = \tilde{\alpha}_k \, \int d^4x\sqrt{g} \, R \, \, \bar{\psi} \gamma^5 \psi \, ,
\ee
which is singled out by this criterion. This term has the structure of a mass term where the mass is set by the spacetime curvature. Upon including $\Delta\Gamma_k^{\rm ferm}$ 
 the flow of $G_k$ and $\Lambda_k$ again takes the form \eqref{betas} with $B^{\rm f,minimal}$ replaced by
\be\label{Bferm}
B^{\rm ferm} = -\frac{1}{12\pi} \left[ 24\alpha + \Big(\pi - 2 \Big) \Big(1 - 12 (\beta + \frac{1}{4}) \Big) \right] \, , 
\ee
where $\alpha_k \equiv \tilde{\alpha}_k \, k$ is the dimensionless counterpart of $\tilde{\alpha}_k$. Eq.\ \eqref{Bferm} indicates that $\alpha$ can play a crucial role in understanding the fixed point structure of the system. In particular, it has the potential to shift all NGFPs to $g_* > 0$ provided that $\alpha_*$ is sufficiently positive to compensate for the regulator contributions. Notably, RG flows including couplings of fermion-bilinears to the spacetime curvature have also been considered in \cite{Eichhorn:2016vvy,Eichhorn:2018nda}.

\section{RG flows including the fermion-curvature coupling}
\label{sect.3}
In sect.\ \ref{sect.2c} we argued that the fermion-curvature coupling \eqref{fermextra} could be essential for understanding the fixed-point structure of gravity-fermion systems. In this section we  complete the analysis by computing the beta function for the coupling $\alpha_k$ (sect.\ \ref{sect.31}) and the analysis of the resulting fixed-point structure in sects.\ \ref{sect.32} and \ref{sect.33}, respectively.
\subsection{Beta functions}
\label{sect.31}
When computing the beta function for $\alpha$, we again make use of the background field method, also retaining the structure 
\be\label{projectionstructure}
\int d^4x\sqrt{\gb}\,\Rb\,\bar{\theta}\, \gamma^5 \, \theta
\ee
on both sides of the projected Wetterich equation. Again we select the background metric to be the one of the four-sphere. Tracking the fermionic terms then also requires adopting a non-zero value of the background fermions $\theta$, c.f.\ eq.\ \eqref{background}. In principle, eq.\ \eqref{projectionstructure} suggests to use covariantly constant spinor fields. On the background sphere this is inconsistent, however, since the Dirac operator does not possess zero modes \cite{Camporesi:1995fb}. We then take $\theta$ as the lowest eigenmode of the Dirac operator, which  satisfies the (generalized) eigenvalue equation
%
$\nabla_\mu \, \theta = i\sqrt{\frac{\Rb}{48}} \, \gamma_\mu \, \theta$.
%

\begin{figure}
	\centering
\includegraphics[width = 0.2\textwidth]{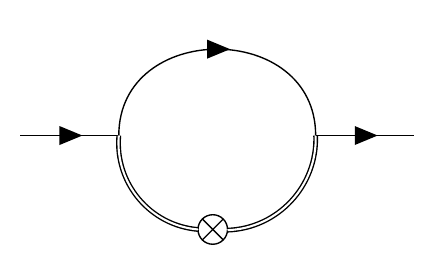}	
\includegraphics[width = 0.2\textwidth]{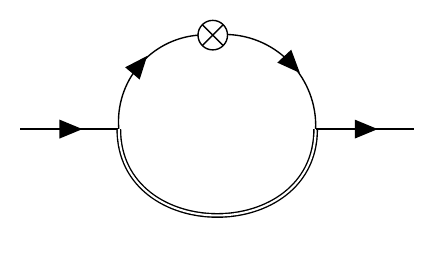} \\ 
\includegraphics[width = 0.17\textwidth]{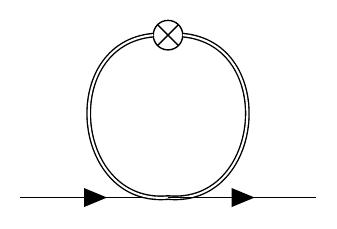}
\caption{\label{Fig.3} Feynman diagrams contributing to $\beta_\alpha$. The bold, external lines correspond to background fermions while the solid single and double lines in the loop encode the propagators of the fermionic fluctuations and graviton fluctuations, respectively. The crossed circle marks the insertion of the regulator $\cR_k$. The three- and four-point vertices are obtained by expanding $\Gamma_k$ about the background field configuration.}
\end{figure}

The beta function
\be\label{betaalpha}
k \p_k \alpha_k = \beta_\alpha(\lambda_k,g_k,\alpha_k)
\ee
 is then found by expanding the right-hand side of the Wetterich equation to second order in the background fermions and identifying the term proportional to $\Rb$. Pictorially, the  contributions to $\beta_\alpha$ are given by the self-energy corrections to the background fermions encoded in 
 the Feynman diagrams shown in Fig.\ \ref{Fig.3}.
Following \cite{Benedetti:2010nr}, the explicit expressions represented by these diagrams are obtained by first splitting $\Gamma_k^{(2)} + \mathcal{R}_k = A+B$ where $A$ is independent of the background fermions and $B$ consists of all terms containing either $\bar{\theta}, \theta$ or both. The entries of $A^{-1}$ are the propagators of the fluctuation fields while $B$ encodes the vertices coupling the fluctuations to the background fields. The inverse $(\Gamma_k^{(2)} + \mathcal{R}_k)^{-1}$ is then constructed as an expansion in $B$:
$\big(A + B\big)^{-1} = A^{-1} - A^{-1}\,B\,A^{-1} + A^{-1}\,B\,A^{-1}\,B\,A^{-1} + \mathcal{O}(B^3)$.
The term $A^{-1}$ is independent of the background fermion field while the terms of order $B^3$ and higher contain at least three powers of the fermion background fields. Thus these terms do not contribute to $\beta_\alpha$. The tad-pole diagram shown in Fig.\ \ref{Fig.3} is then generated by the term of order $B$ while the diagrams containing the three-point vertices arise at order $B^2$. 
Thus Fig.\ \ref{Fig.3} then includes all diagrams that contribute to the self-energy of the background fermions. 

The Feynman diagrams imply that $\beta_\alpha$ will be a polynomial of degree three in $\alpha$.
The presence of the cubic term is inferred from the observation that each three-point vertex contains a term that is proportional to $\alpha_k$. In addition the fermion propagator contains a term proportional to $\alpha_k \Rb$. The projection onto \eqref{projectionstructure} then entails an expansion of the fermion propagator in the background curvature, so that the first two diagrams in Fig.\ \ref{Fig.3} contain contributions up to order $\alpha^3$. Hence
\be\label{betaalphapara}
\beta_\alpha = A_0 + (A_1+1)\,\alpha + A_2\,\alpha^2 + A_3\,\alpha^3 \, . 
\ee
The coefficients $A_i$, $i=0,1,2,3$, depend on the couplings $\lambda, g$ as well as the coarse-graining parameter $\beta$,
\be\label{Aipara}
A_i(\lambda,g) = \frac{g}{\pi} \left( \frac{A_i^1}{(1-2\lambda)} + \frac{A_i^2}{(1-2\lambda)^2} + \frac{A_i^3}{(1-2\lambda)^3} \right) \, ,
\ee
with the non-zero numerical coefficients $A_i^j$, $j = 1,2,3$, being
\begin{equation}
\begin{split}
    &A_0^1 = -\frac{3}{32}\\
    &A_0^2 = \frac{3}{8} - \frac{15\pi}{128} + \frac{1}{32}\eta_N\\    
     &A_0^3 = \frac{7}{20}-\frac{3\pi}{32} - (\frac{179}{1120} - \frac{3\pi}{64})\eta_N \\
     &A_1^1 = -\frac{7}{6} + \frac{7\pi}{16} - \frac{\beta}{2}\\
     &A_1^2 = \frac{107}{30} + \frac{\pi}{32} + (\frac{1}{30}-\frac{13\pi}{64})\eta_N
     -\frac{\beta}{4} - (\frac{39}{40}-\frac{21\pi}{64})\beta\eta_N\\
     &A_1^3 = -\frac{67}{30} -\frac{\pi}{8} + (\frac{47}{210} + \frac{\pi}{32})\eta_N\\
    &A_2^1 = \frac{47}{12} - \frac{5\pi}{4} + (\frac{45}{4} - \frac{45\pi}{16})\beta\\
  &A_2^2 = \frac{169}{120} - \frac{\pi}{2} + (\frac{101}{280}-\frac{\pi}{8})\eta_N\\ &\qquad \qquad + (\frac{9}{2}-\frac{9\pi}{8})\beta - (\frac{61}{20} - \frac{15\pi}{16})\beta\eta_N\\
  &A_2^3 = -\frac{17}{105}+(\frac{143}{630}-\frac{\pi}{16})\eta_N\\
    &A_3^1 = \frac{9}{10}\\
    &A_3^2 = -\frac{17}{10} + (\frac{79}{28}-\frac{27\pi}{32})\eta_N\\
 \end{split}
\end{equation}
The explicit form of the beta function \eqref{betaalphapara} constitutes the main result of this subsection.

\subsection{Fixed-point structure of the extended system}
\label{sect.32}
When investigating the fixed point structure for the $\lambda$-$g$-$\alpha$ system, we start with the following observations:
\begin{enumerate}
\item[a)] Including the fermionic sector $\Gamma_k^{\rm ferm} + \Delta\Gamma_k^{\rm ferm}$ supplements the beta functions \eqref{betas} with the beta function \eqref{betaalphapara}. Notably, $\beta_\alpha$ is cubic in $\alpha$. This guarantees that, for fixed $\lambda, g$, the equation $\beta_\alpha = 0$ has at least one real solution. Thus the NGFPs seen in the approximation $\Delta\Gamma_k^{\rm ferm} = 0$ will persist once the fermion-curvature coupling is added.
\item[b)] The coefficient $A_0$ in $\beta_\alpha$ is non-zero. As a consequence, the NGFPs found at minimal coupling do not generalize to fixed points with $\alpha_* = 0$. Quantum gravity fluctuations shift the position to a non-zero value $\alpha_*$. This shift is generated by the fermion-kinetic term and becomes visible once $\Delta\Gamma_k^{\rm ferm}$ is included. This mechanism is identical to the one generating the gravity-induced non-vanishing scalar couplings \cite{Eichhorn:2012va}.
\item[c)] When investigating the transition from $N_f=0$ to a small value, $N_f = 10^{-3}$ say, one finds that the fixed point seen at minimal coupling branches into 3 families of NGFPs discriminated by their value for $\alpha_*$. In addition there is one family of NGFPs coming in from $\alpha_* \rightarrow - \infty$.
\item[d)] Increasing $N_f$ the NGFPs coming from $\alpha_* \rightarrow - \infty$ annihilate with one branch of NGFPs emanating from the gravitational fixed point. This occurs at $N_f \approx 3$.
\item[e)] Performing a large-$N_f$ expansion of the beta functions \eqref{betaEH} and \eqref{betaalphapara}, one establishes that the two other branches extend to infinite families of NGFPs existing for all values of $N_f$. These solutions will be labeled NGFP$^{\rm A}$ and NGFP$^{\rm B}$. 
\end{enumerate}

%
%
%
%
%
%

Upon exhibiting the general structure of the fixed point system, we now investigate the properties of the solutions NGFP$^{\rm A}$ and NGFP$^{\rm B}$ numerically. 
\begin{figure*}[t!]
	\centering
	\includegraphics[width = 0.4\textwidth]{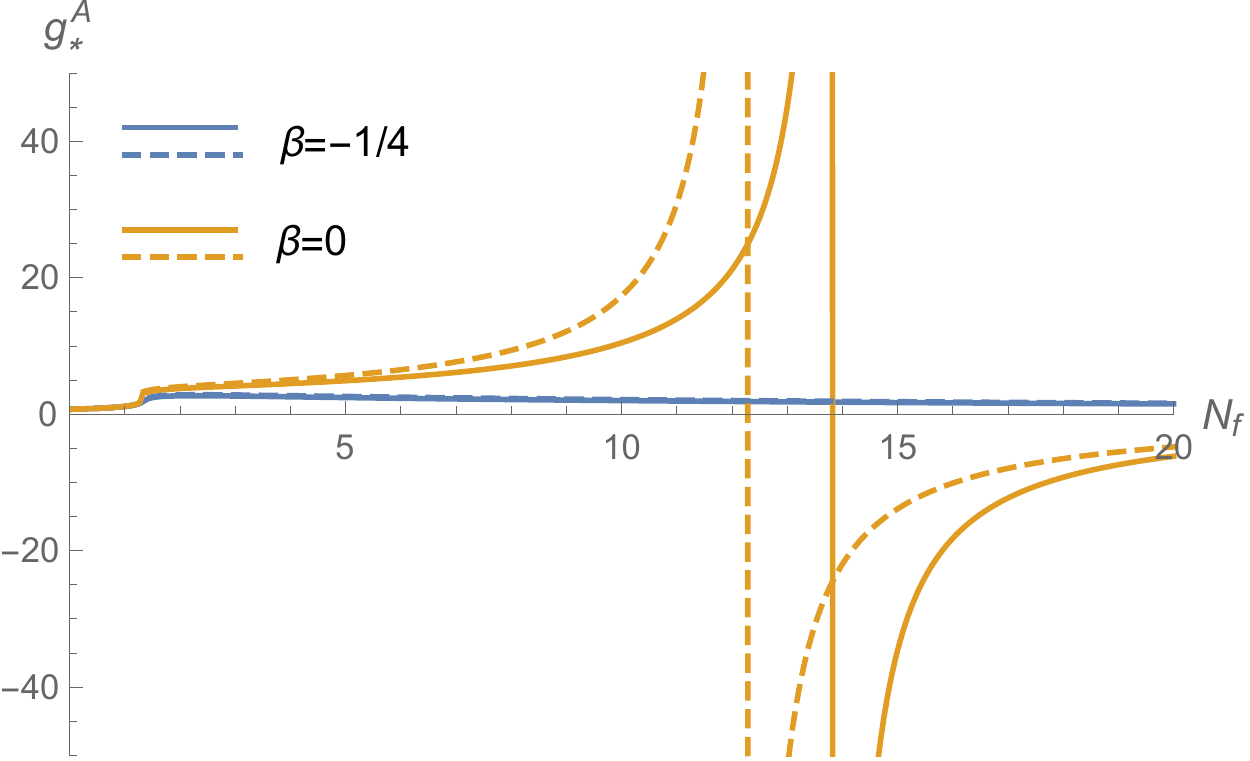}	\, \,
	\includegraphics[width = 0.4\textwidth]{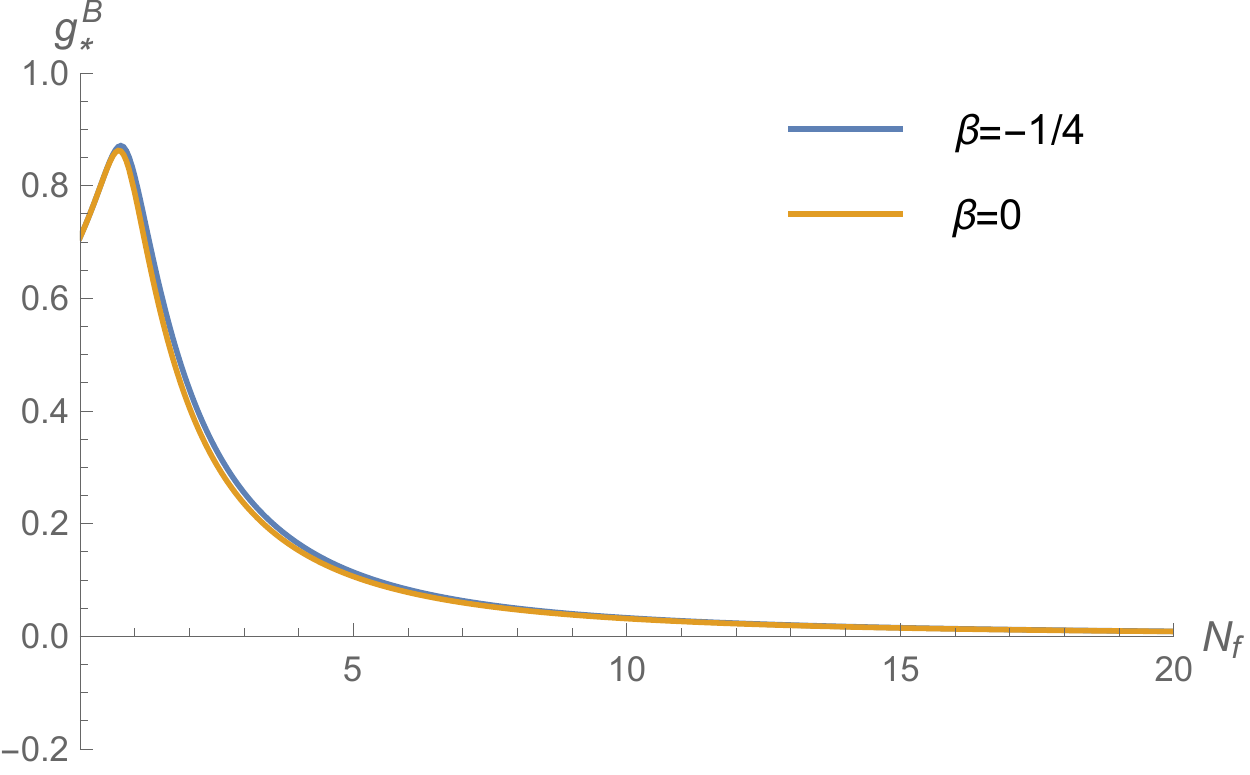} \\[2ex]
	\includegraphics[width = 0.4\textwidth]{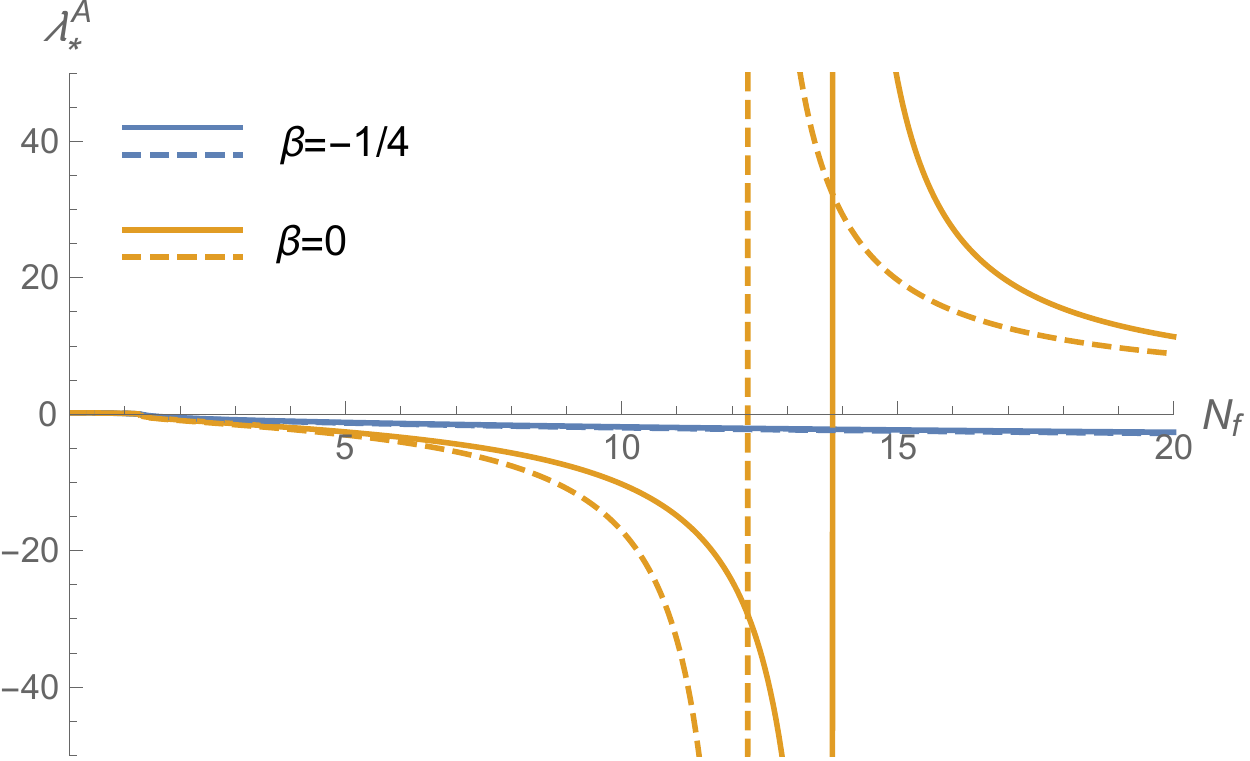}	\, 
	\includegraphics[width = 0.4\textwidth]{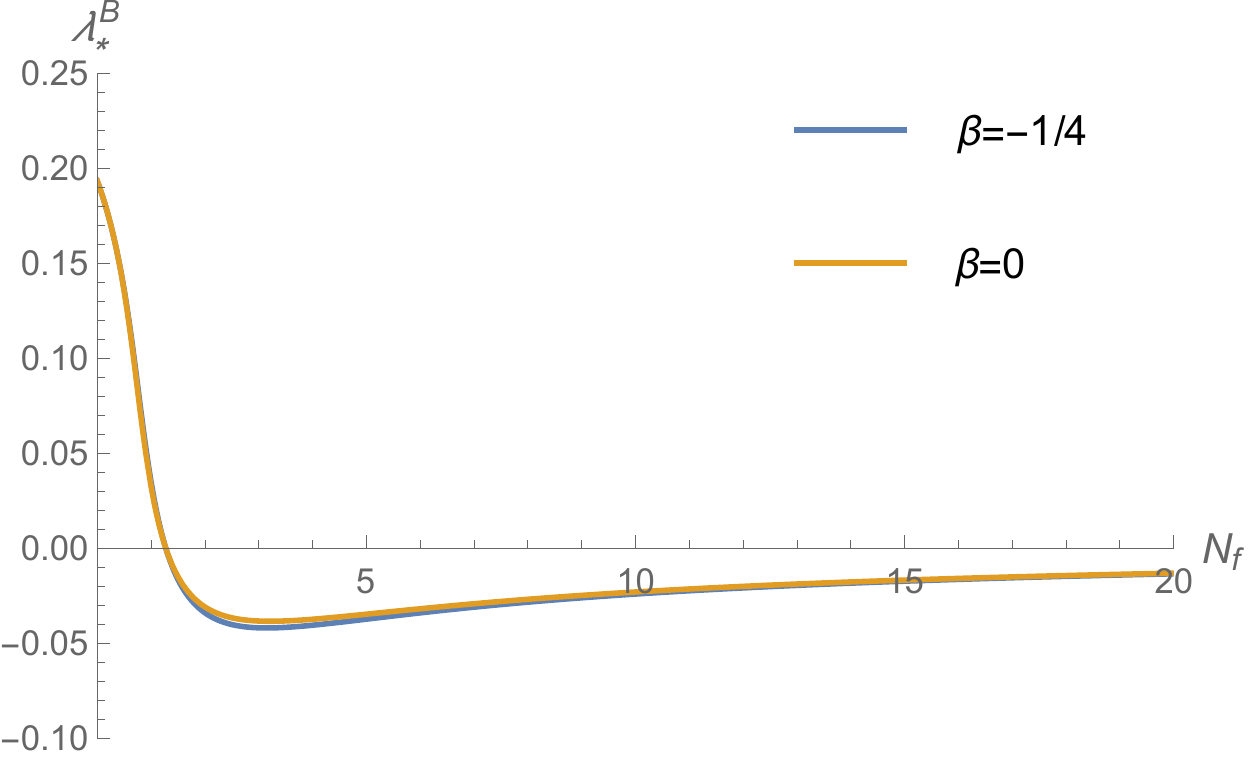} \\[2ex]
	\includegraphics[width = 0.4\textwidth]{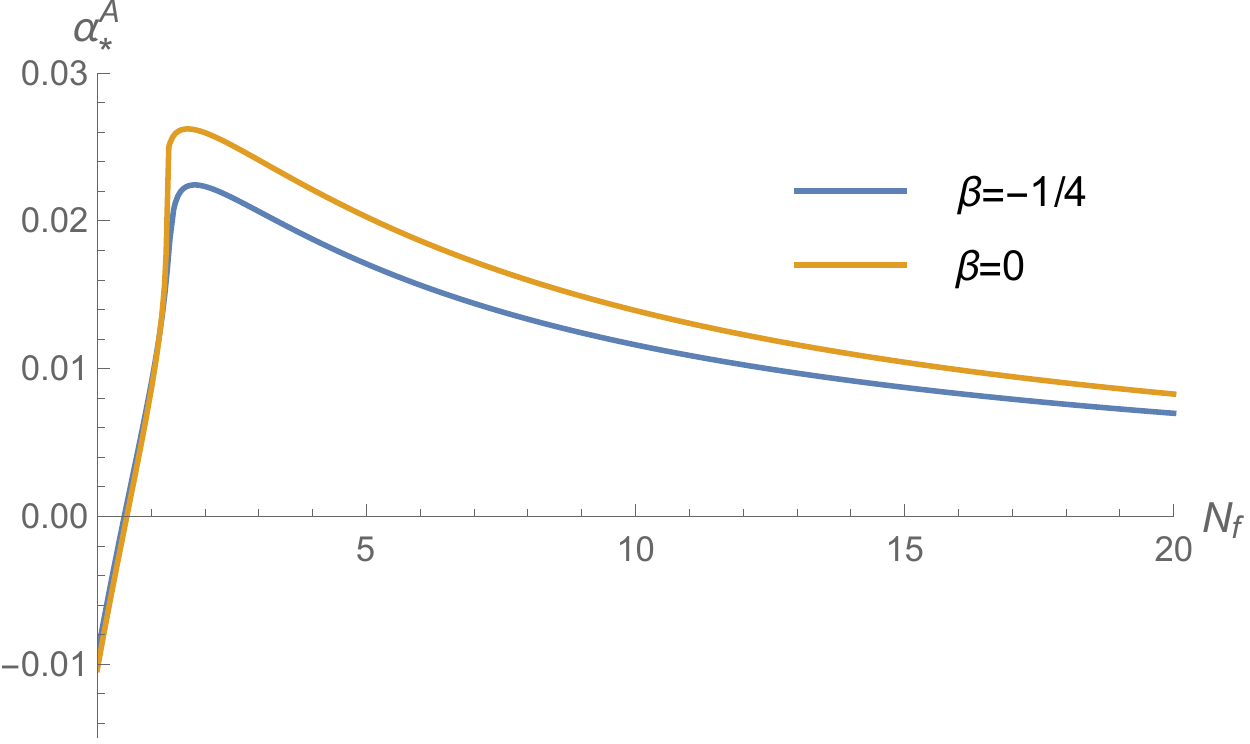}	\, \,
	\includegraphics[width = 0.4\textwidth]{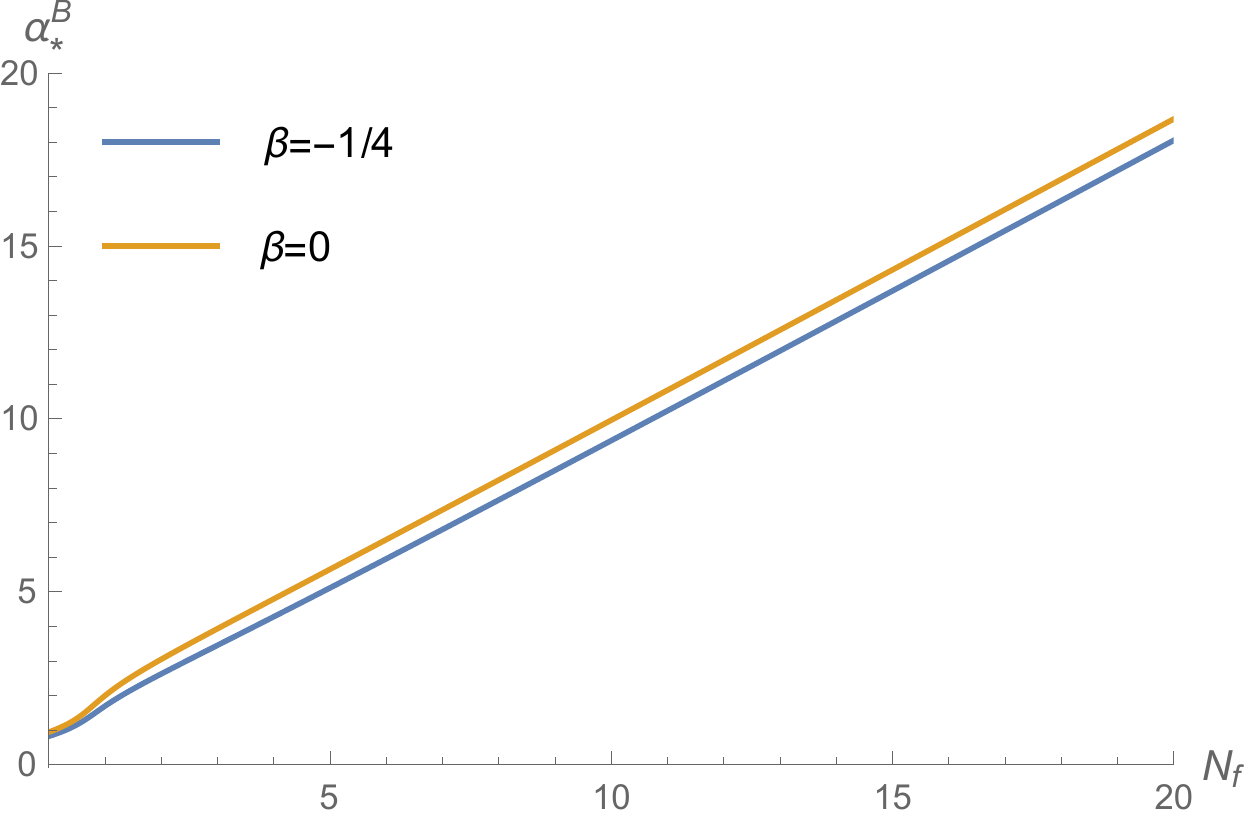}
	\caption{\label{Fig.2} Position of the fixed point solutions NGFP$^{\rm A}$ and NGFP$^{\rm B}$ arising from the extended beta functions \eqref{betaEH} and \eqref{betaalphapara} as a function of $N_f$. The blue and orange lines correspond to the coarse-graining operator being the Laplacian ($\beta = -1/4$) and the squared Dirac operator ($\beta = 0$), respectively. The dashed lines shown in the diagrams for $g_*^{\rm A}$ and $\lambda_*^{\rm A}$ correspond to the position of the NGFP found at minimal coupling, $\alpha = 0$. For $\beta = 0$, NGFP$^{\rm A}$ is shifted into the unphysical region $g_* < 0$ when $N_f > N_f^{\rm crit}$.
}
\end{figure*}
Their position $\{g_*, \lambda_*, \alpha_* \}$ as a function of $N_f$ is shown in Fig.\ \ref{Fig.2}. The analysis establishes the value $\alpha_*$ as the feature distinguishing the two branches: family A is characterized by $\alpha_*^{\rm A} \ll 1$ with $\alpha_*^{\rm A}$ decreasing for increasing $N_f$. For family B, $\alpha_*^{\rm B} \propto N_f$ increases linearly with the number of fermions. The comparison between the solid and dashed lines shows that NGFP$^{\rm A}$ actually shares all the essential properties of the NGFPs found in the minimally coupled case. In particular, the position $g_*^{\rm A}$ is again sensitive to the choice of $\beta$: for $\beta = 0$ there is a critical number of fermions $N_f^{\rm crit}$ at which the solution transits from $g_*^{\rm A} > 0$ ($N_f < N_f^{\rm crit})$ to $g_*^{\rm A} < 0$ ($N_f > N_f^{\rm crit})$. The similarity of NGFP$^{\rm A}$ to the minimally coupled case can be understood easily from the fact that $\alpha_*^{\rm A} \ll 1$ so that $\alpha_* = 0$ constitutes a good approximation.

The family NGFP$^{\rm B}$ is situated at $g_*^{\rm B} > 0$ for all values of $N_f$, i.e, there is no critical number of fermions independent of the choice of coarse-graining operator. This feature can be understood by revisiting the fixed point condition $\eta_N^* = -2$: since the value $\alpha_*$ increases with an increasing number of fermions, the contribution of the coupling $\alpha$ always dominates over the contribution of the regulator. With $\alpha_* > 0$ one then finds that the solution of $\eta_N^* = -2$ is always situated at $g_* > 0$. Moreover, $\alpha_* \gg \beta$ also ensures that the position of the fixed point is largely independent of $\beta$.

Finally, one can study the stability of the RG flow in the vicinity of the two families of fixed points by computing the eigenvalues of the stability matrix $\bf B$, see Table \ref{Tab.1} for typical examples. This reveals that the fixed points NGFP$^{\rm A}$ and NGFP$^{\rm B}$ come with two and three relevant directions, respectively. This result is independent of $N_f$ and the choice of $\beta$. As a consequence, a high-energy completion based on a NGFP from the family A may predict the low-energy value of $\alpha$ while for the fixed points in the family B this value corresponds to a free parameter which must be taken from experiment. 

\begin{table}
	\begin{tabular}{cccccc}
fixed point & \; \; $N_f$ \; \; & \; \; $\beta$ \; \; & \; $\theta_1$  \; &  \; $\theta_2$  \; &  \; $\theta_3$  \; \\ \hline
\multirow{4}{*}{NGFP$^{\rm A}$} 
& $1$ & $0$ & \multicolumn{2}{c}{$0.72\pm 2.04 i$} & $- 1.48$\\
& $1$ & $-1/4$ &  \multicolumn{2}{c}{$0.84\pm 1.90 i$} & $- 1.62$  \\
& $20$ & $0$ & $4.03$ & $2.19$ & $-1.00$ \\
& $20$ & $-1/4$ & $3.92$ & $2.04$ & $-1.08$ \\ \hline
\multirow{4}{*}{NGFP$^{\rm B}$} & $1$ & $0$ & \multicolumn{2}{c}{$2.79\pm 1.16 i$} & $1.05$ \\
& $1$ & $-1/4$ &  \multicolumn{2}{c}{$2.56\pm 1.16 i$} & $1.09$ \\
& $20$ & $0$ & $3.91$ & $3.79$ & $0.57$ \\
& $20$ & $-1/4$ & $3.93$ & $3.73$ & $0.56$ \\ \hline
\end{tabular}
\caption{\label{Tab.1} Typical values for the critical exponents associated with the two families of NGFPs. Notably, the $\beta$-dependence of the critical exponents is rather minor, showing that the physical properties of the fixed points are actually robust with respect to changing the coarse-graining scheme.}
\end{table}

\subsection{Flow diagram and predictivity}
\label{sect.33}
We close this section by illustrating the RG flow created by the beta functions \eqref{betaEH} and \eqref{betaalphapara}. For this purpose, we project the full system onto the $\alpha$-$g$--plane by setting $\lambda = 0$.  For concreteness, we choose $N_f = 3$ and $\beta = -1/4$ which serves as an illustrative example of the general situation where one has two NGFPs situated at $g_* > 0$. The flow is then controlled by the projection of the fixed points found for the full $\lambda$-$g$-$\alpha$-system:
$\{\alpha_*^{\rm GFP}, g_*^{\rm GFP}\} = \{0,0\}$, 
$\{\alpha_*^{\rm A}, g_*^{\rm A}\} = \{0.02,1.49\}$, and
$\{\alpha_*^{\rm B}, g_*^{\rm B}\} = \{2.82,0.30\}$. 
The GFP serves as an infrared attractor, capturing the RG flow in its vicinity as $k$ is lowered. NGFP$^{\rm A}$ is a saddle point with the UV-attractive direction almost aligned with the $\alpha=0$-axis. The NGFP$^{\rm B}$ is UV-attractive in both $\alpha$ and $g$. The stability properties of NGFP$^{\rm B}$ are remarkable in the sense that the two right-eigenvectors of the projected stability matrix are almost parallel, enclosing an opening angle $\theta \simeq 13^\circ$.

\begin{figure}[t!]
	\centering
	\includegraphics[width = 0.48\textwidth]{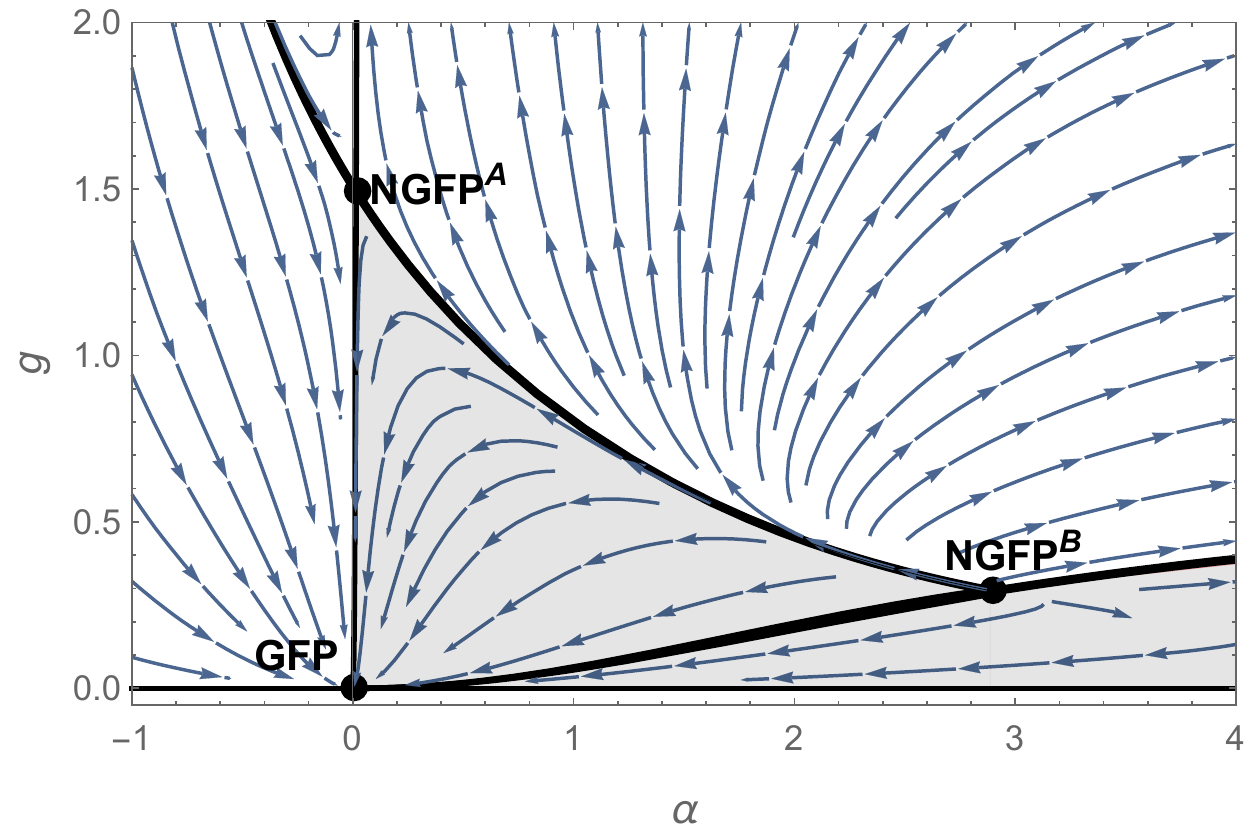}
	\caption{\label{Fig.4} Illustration of the RG flow projected onto the $\alpha$-$g$--plane obtained for $N_f = 3$ and $\beta = -1/4$. The arrows point towards the infrared, i.e., smaller values of $k$. The positions of the projected fixed points are marked by black dots. The thick black line marks the boundary of the region where the flow is attracted to the GFP at low energy. The figure provides a prototypical example of the interplay between the fixed points.}
\end{figure}
The flow generated by the projected beta functions is shown in Fig.\ \ref{Fig.4}. The black lines originate from integrating the beta functions with initial conditions set along the eigenvectors of the stability matrices associated with NGFP$^{\rm A}$ and NGFP$^{\rm B}$. The bold black line connecting the NGFPs acts as a boundary: RG trajectories below this line are attracted to the GFP as $k \rightarrow 0$ while trajectories above this line typically terminate at a finite value of $k$. The trajectories in the shaded region are complete in the sense that their flow interpolates between the NGFP$^{\rm B}$ for $k \rightarrow \infty$ and the GFP as $k \rightarrow 0$. The approach to the GFP then guarantees the existence of a classical low-energy regime where general relativity constitutes a good approximation of the gravitational physics.

\section{Conclusions}
\label{sect.4}
Motivated by the asymptotic safety scenario \cite{Percacci:2017fkn,Reuter:2019byg} for gravity-matter systems, we studied the fixed point structure of gravity coupled to $N_f$ Dirac fermions on a spherically symmetric background. Driven by the significant regulator dependence found at minimal coupling \cite{Dona:2012am}, our work extended the minimal case by adding a distinguished coupling between the fermion bilinears and the Ricci scalar constructed from the spacetime metric. This coupling is induced dynamically by quantum gravity fluctuations. As a main result, we identified the two infinite families of non-Gaussian renormalization group fixed points (NGFPs) shown in Fig.\ \ref{Fig.2} and existing for all values $N_f$. The first family shows the behavior reported in \cite{Dona:2012am}, possibly exhibiting an upper critical number of fermions for which the fixed points could be used in asymptotic safety. The second one could provide a phenomenologically viable high-energy completion for all values of $N_f$. We stress that both families of fixed points could provide UV-completions of phenomenologically interesting gravity-matter systems once suitable bosonic degrees of freedom are added. Table \ref{Tab.1} thereby suggests that the family NGFP$^{\rm A}$ comes with an enhanced predictive power as compared to NGFP$^{\rm B}$.

In our work, we specifically analyzed the effect of implementing different coarse-graining schemes \eqref{boxdef} based on the squared Dirac operator ($\beta=0$) and Laplacian $(\beta = -1/4)$, respectively. In combination with the fermion-curvature coupling $\alpha$ introduced in eq.\ \eqref{fermextra} this resulted in a rather clear picture: quantum gravity fluctuations dictate that $\alpha$ must be non-zero at any NGFP. For the first family of NGFPs, $\alpha_*$ is very small and well-approximated by $\alpha_* \simeq 0$. The regularization procedure based on the squared Dirac operator then induces contributions to the flow which dominate over this coupling. As a result, the position of the NGFPs exhibits a ``strong'' dependence on the choice of coarse-graining operator. Notably, neither the existence nor the critical exponents are sensitive to the choice of $\Box$. It is merely the shift in position which may render the NGFPs unsuitable for a phenomenologically valid high-energy completion of the gravity-fermion system. Conversely, the second family of fixed points comes with much larger values $\alpha_*$. As a consequence, the contribution from the regulator is subdominant and the fixed point properties show only a minor dependence on $\Box$. This  suggests that it is actually the Laplacian which is the canonical choice for the coarse-graining operator, since other values of $\beta$ induce additional interaction terms originating from the regularization procedure.  

At this stage, it is useful to compare our findings to previous studies of gravity-fermion systems. Fig.\ \ref{Fig.2} shows that our results are completely in line with \cite{Dona:2012am}. The solutions NGFP$^{\rm B}$ go unnoticed in this study since the curvature-fermion coupling distinguishing the two families of NGFPs is not included. Comparing to  \cite{Eichhorn:2016vvy,Eichhorn:2018nda}, the stability properties reported in \cite{Eichhorn:2016vvy} suggest that the NGFP$^{\rm A}$ could be the ``chiral non-Gaussian'' fixed point while NGFP$^{\rm B}$ shares the stability properties of the ``non-Gaussian'' fixed point identified in their Table I. Moreover, the family NGFP$^{\rm A}$ (with $\beta = 0$) exhibits the same structure as the background couplings reconstructed from the fluctuation computation \cite{Meibohm:2015twa}, suggesting that the computations actually probe the same universality class. In this context, it has also been argued that it may not be the Newton’s coupling $g_k$ analyzed in Fig. \ref{Fig.2} which governs the gravitational interactions with matter and thus has to satisfy the positivity bound $g^* \ge 0$. This interpretation opens the possibility that both families NGFP$^{\rm A}$ and NGFP$^{\rm B}$ provide phenomenologically interesting UV-completions of gravity-matter systems for all values of $N_f$. While the computation of the fluctuation coupling analyzed in \cite{Meibohm:2015twa,Eichhorn:2018nda} is beyond the scope of this letter, it would be very interesting to investigate if the existence of the family NGFP$^{\rm B}$ can be corroborated within the fluctuation approach. The interaction \eqref{projectionstructure} could thereby serve as a guiding principle towards the tensor structures which should be studied in this context.

\section*{Acknowledgements}
\noindent
We thank A.\ Vereijken for fruitful discussions and A.\ Eichhorn and R.\ Percacci for useful comments on the manuscript. J.W.\ acknowledges the China Scholarship Council (CSC) for financial support.



\end{document}